\documentclass[10pt,nofootinbib,notitlepage,superscriptaddress,twocolumn]{revtex4-1}
\pdfoutput=1

\usepackage[utf8]{inputenc} 		
\usepackage[ngerman,english]{babel} 
\usepackage[T1]{fontenc}    		
\usepackage{outlines}
\usepackage{amsmath,amssymb}
\usepackage{feynmp}
\usepackage{graphicx}
\usepackage{slashed}
\usepackage[small, bf, flushleft]{caption}
\usepackage[caption=false]{subfig}

\usepackage{hyperref}
\hypersetup{
    colorlinks,
    citecolor=black,
    filecolor=black,
    linkcolor=blue,
    urlcolor=blue
}

\newcommand{\eV}{\,\mathrm{eV}}
\newcommand{\km}{\,\mathrm{km}}

\begin{document}
\title{Decoherence of Gravitational Wave Oscillations in Bigravity}

\author{Kevin Max}\email{kevin.max@sns.it}
\affiliation{Scuola Normale Superiore and INFN Pisa,\\ 
Piazza dei Cavalieri, 7 - 56126 Pisa, Italy}
\author{Moritz Platscher}\email{moritz.platscher@mpi-hd.mpg.de}
\affiliation{Max-Planck-Institut f\"ur Kernphysik, \\
Saupfercheckweg 1, 69117 Heidelberg, Germany}
\author{Juri Smirnov}\email{juri.smirnov@fi.infn.it}
\affiliation{INFN divisione di Firenze, Dipartimento di Fisica, Università di Firenze,\\
Via Sansone 1, 50019 Sesto Fiorentino, Florence, Italy}

\begin{abstract}
\begin{center}
\textbf{ABSTRACT}
\end{center}

\noindent Following up on our recent study, we consider the regime of graviton masses and gravitational wave propagation distances at which decoherence of the wave packets plays a major role for phenomenology. This regime is of particular interest, as it can lead to very striking phenomena of echo events in the gravitational waves coming from coalescence events. The power of the experimental search in this case lies in the fact that it becomes sensitive to a large range of graviton masses, while not relying on a specific production mechanism. We are thus able to place new relevant limits on the parameter space of the graviton mixing angle.

\end{abstract}

\maketitle

\section{Introduction}

In a recent publication~\cite{PhysRevLett.119.111101} we have pointed out the possibility that gravitational waves (GWs) can oscillate in close analogy to neutrinos if the mediator of the gravitational force has a mass. This effect arises in the framework of bigravity, as to the present day it turns out to be the only consistent framework for massive gravity in four space-time dimensions.~\cite{deRham:2010ik,deRham:2010kj,deRham:2011rn,Hassan:2011hr,Hassan:2011vm,Hassan:2011tf,Hassan:2011zd,Hassan:2011ea,Comelli:2012vz,Deffayet:2012nr,Deffayet:2012zc} The effect is linked to the fact that in bigravity, there exists a tensor field which couples directly to matter as a physical metric, just as in general relativity (GR), and a \emph{sterile} tensor field acting to non-linearly implement the St\"uckelberg mechanism~\cite{Stueckelberg:1900zz}. The additional, sterile tensor field is only coupled to the physical metric via a potential term and in the linear regime a close analogy to sterile neutrinos can be observed, see also~\cite{Berezhiani:2007zf,Hassan:2012wr,DeFelice2014,Narikawa:2014fua,Brax:2017hxh} for previous studies.

An important aspect of massive gravity, which is important to highlight at this point, is the (conjectured) Vainshtein mechanism~\cite{Vainshtein:1972sx}. In the allowed graviton mass range, all astrophysical processes such as solar system observations, binary coalescence, and others take place \emph{inside} the so-called Vainshtein sphere; e.g.~for the masses involved in the binary black hole (BBH) merger event GW150914, the Vainshtein radius \cite{Babichev:2013pfa} is $r_V\approx 8\times10^{11}\text{ m}$ for $m_g=10^{-22}\,$eV, much larger than the interaction distance of the merging BHs ($\sim100\dots1000\km$) and their Schwarzschild radii ($\sim10\dots 100\km$). In this sphere the longitudinal graviton mode is strongly coupled and the system behaves as in pure GR. This implies that in a merger event, GWs are produced exactly as in GR, but in bigravity only the linear combination which couples to matter is produced. As in neutrino physics this linear combination is a superposition of two mass eigenstates, which is coherent owing to the Vainshtein mechanism at~production.

When a GW propagates through space, the effect discussed in \cite{PhysRevLett.119.111101} takes place while the waves are still coherent: as long as the condition $d_L \approx L_\text{coh}$ is satisfied [where $d_L$ is the luminosity distance and the coherence length $L_\text{coh}$ is defined in Eq.~\eqref{eq:coherence_length}], the oscillation can have a detectable effect on the GW signal shape and thus is distinguishable from GR. For the details of the modified shape analysis, we refer to~\cite{PhysRevLett.119.111101}. If, however, $d_L>L_\text{coh}$, the~wave packets decohere, i.e.~the spatial distance between the propagating mass eigenstates is larger than the sizes of the corresponding wave packets. Consequently, a GW detector sees only an overall reduction of the strain compared to GR, and a second signal may appear. This effect is the main purpose of this publication: We discuss how it can be used to falsify or verify the existence of GW oscillations on the basis of a (large) number of BBH merger observations.

With current observations, we show that we are able to probe the parameter range $m_g \gtrsim 10^{-22}\,$eV and small mixing angle (as defined in Sec.~\ref{subsec:bimetric_action}). For a comprehensive reference of graviton mass studies, see~\cite{deRham:2016nuf}. Therein, the most stringent model-independent bound is $m_{\text{bound}}~\leq~7.2~\times~10^{-23} \eV$, found from solar system tests. Note however that this probes only the pure massive gravity case, i.e.~bigravity with $\theta=\pi/2$.

This article is structured as follows. In Sec.~\ref{sec:GWoscillations} we review the basic concepts of GW oscillations, giving a slightly different derivation than in~\cite{PhysRevLett.119.111101} with canonically normalised states; however, the results are consistent. A detailed discussion of the observable effects is given in Sec.~\ref{sec:results}, where we analyse the modified BBH merger rate and possible echo signals. We summarise our results in Sec.~\ref{sec:Conclusions} and put them into perspective with complementary studies.


\section{Summary of Gravitational wave oscillations} \label{sec:GWoscillations}

Before turning our attention to GW signals in bigravity, we establish the framework and conventions used in this article.
\subsection{The bimetric action}
\label{subsec:bimetric_action}

The fully nonlinear gauge invariant, ghost-free action of two dynamical metrics is~\cite{Hassan:2011zd},
\begin{widetext}
\begin{equation}
\begin{split}
          S_{\text{Bi}} =&\, \frac{M_g^2}{2} \int d^4 x \sqrt{- \det g} \ R(g)  +\frac{M^2_{\tilde{g}}}{2} \int d^4 x \sqrt{- \det \tilde{g}} \ \tilde{R}(\tilde{g}) \, +\\
          &+ m^2 M^2_\text{eff} \int d^4 x  \sqrt{-\det g}  \sum_{n=0}^4 \beta_n e_n(\sqrt{g^{-1} \tilde{g}}) +  \int d^4 x  \sqrt{- \det g}  \,\,\mathcal{L}_{\text{matter}}\,,
    \label{eq:bigrav_action}
\end{split}
\end{equation}
\end{widetext}
where $M_g$ ($M_{\tilde{g}}$) is the Planck mass of the physical metric  $g$ (sterile tensor field $\tilde{g}$), $M_{\text{eff}}^{-2} \equiv M_g^{-2} + M_{\tilde{g}}^{-2}$. The kinetic terms of the tensors are given by the corresponding Ricci scalars $R(g)$ and $\tilde{R}(\tilde{g})$. The potential terms which link the two tensors $g$ and $\tilde{g}$ are given by the constants $\beta_n$ and matrix polynomials of $\mathbb{X}\equiv \sqrt{g^{-1} \tilde{g}}$ ($[\mathbb{X}] \equiv \mathrm{tr} \mathbb{X}$):
\begin{equation}
\begin{gathered}
e_0(\mathbb{X})=1\, \ e_1(\mathbb{X})=\left[\mathbb{X} \right], \ e_2(\mathbb{X})=\frac{1}{2}\left(\left[\mathbb{X} \right]^2-\left[\mathbb{X}^2 \right]\right)\,,\\
e_3(\mathbb{X})=\frac{1}{6}\left(\left[\mathbb{X} \right]^3-3\left[\mathbb{X} \right]\left[\mathbb{X}^2 \right]+2\left[\mathbb{X}^3 \right]\right)\,, \\
e_4(\mathbb{X})=\det \mathbb{X}\,.
\end{gathered}
\end{equation}
We also define the mixing angle $\sin^2(\theta)=M_{\text{eff}}^2/M_g^2$, $\cos^2(\theta)=M_{\text{eff}}^2/M_{\tilde{g}}^2$. Note from \eqref{eq:bigrav_action} that we assume the matter sector to only couple to the metric $g$, a requirement necessary for a consistent theory valid up to high scales.~\cite{deRham:2014fha} In this setup, standard GR plus matter is recovered in the limit $\theta \rightarrow 0$. Note that the (in)famous vDVZ discontinuity in the limit $m \to 0$ is avoided in this limit, which can be considered the decoupling limit of the sterile tensor field.~\cite{vanDam:1970vg,Zakharov:1970cc} 

We perturb both metrics about an FRW-like background with conformal time $\eta$,~\cite{vonStrauss:2011mq,Comelli:2011zm}
\begin{equation}
\begin{split}
\mathrm{d}s^2 \equiv g_{\mu\nu} \mathrm{d}x^\mu  \mathrm{d}x^\nu =&\, a(\eta)^2(-\mathrm{d}\eta^2 + \mathrm{d}\vec{x}^2)\,, \\ \mathrm{d}\tilde{s}^2 \equiv \tilde{g}_{\mu\nu} \mathrm{d}x^\mu  \mathrm{d}x^\nu  =&\, b(\eta)^2(-\tilde{c}(\eta)^2\,\mathrm{d}\eta^2 + \mathrm{d}\vec{x}^2)\,,
\end{split}
\end{equation}
by splitting the tensors into the background and a (small) perturbation,~\cite{Comelli:2012db}
\begin{equation}
\begin{split}
g_{\mu\nu} &= a^2(\eta)\left( \eta_{\mu\nu} +\frac{h_{\mu\nu}(x,\eta)}{M_g} \right),\\
\tilde{g}_{\mu\nu} &= b^2(\eta)\left( \eta_{\mu\nu} +\frac{\tilde{h}_{\mu\nu}(x,\eta)}{M_{\tilde{g}}} \right).
\end{split}
\end{equation}
Note that the lapse of the reference background metric can safely be set to one, as we are interested in late time solutions (see~\cite{PhysRevLett.119.111101} for a derivation). In this limit, a static cosmological background solution implies that the ratio of scale factors $y\equiv b/a$ approaches a constant value $y_*$~\cite{vonStrauss:2011mq} and $c = 1$.

We now choose a transverse traceless gauge for both metrics,\footnote{In fact, the transverse traceless condition is always satisfied for the purely massive mode, since it is a gauge invariant quantity.~\cite{Schmidt-May:2015vnx}} which leaves two helicity-2 excitations for each metric.\footnote{We can ignore the helicity-1 modes (as they do not couple to the energy-momentum tensor) and scalar modes (which are screened due to the Vainshtein effect~\cite{deRham:2014zqa}).} The calculation is furthermore simplified by noting that $a(\eta)=\frac{1}{1+z} = 1~+~\mathcal{O}(.1)~\approx~\text{const}$~for~the~distances of interest, e.g.~of the event GW150914 with $z \approx 0.09$. The potential in~\eqref{eq:bigrav_action}, to quadratic order, reads
\begin{equation}
\begin{split}
          S^{(2)}_{\text{Bi}} \supset \int\,d^4x \frac{m^2 M^2_\text{eff}}{8} a^4 \,y_* \Gamma_* \Big( \frac{\tilde{h}_{\mu\nu}}{M_{\tilde{g}}} -\frac{h_{\mu\nu}}{M_g} \Big)^2  
    \label{eq:quad_potential}
\end{split}
\end{equation}
where we have defined $\Gamma_* \equiv (\beta_1+2y_* \beta_2 + y_*^2 \beta_3) $, which is exactly the combination of parameters which arises in the cosmological solution within bigravity. \cite{PhysRevLett.119.111101,Comelli:2011zm,vonStrauss:2011mq}

\subsection{Tensor mode oscillations} 

The equations of motion of linearised bigravity in transverse traceless gauge are derived from \eqref{eq:quad_potential},~\cite{Comelli:2012db}
\begin{subequations}
\begin{align}
h''+ k^2 h +\frac{m^2}{2}   \,\Gamma_* a^4 y_* \sin \theta \Big(\sin \theta \, h - \cos \theta \, \tilde{h} \Big) &= 0\,,\\
\tilde{h}'' + k^2 \tilde{h} +\frac{m^2}{2}   \,\Gamma_* a^4 y_* \cos \theta \Big(\cos \theta \, \tilde{h} - \sin \theta \, h \Big) &= 0\,,
\end{align}
\end{subequations}
where the polarisation index $h_{\times,+}$ has been omitted, and $k=|\vec{k}|$ denotes the three-momentum.  We diagonalise the equations of motion with the field redefinition\footnote{Note that the redefinition is a simple rotation because the fields $h,\tilde{h}$ have been normalised canonically, different to Refs.~\cite{Comelli:2011zm,PhysRevLett.119.111101}.}
\begin{equation}
\left(
\begin{array}{c}
h_1\\
h_2\\
\end{array}
\right) \equiv
\left(
\begin{array}{cc}
\cos\theta & \sin\theta \\
-\sin\theta & \cos\theta \\
\end{array}
\right)
\left(
\begin{array}{c}
h\\
\tilde{h}\\
\end{array}
\right)
\label{eq:diagonaliseEOM}
\end{equation}
which yields the equations of motion in the mass basis,
\begin{subequations}
\label{eq:eom_mass_eigenstates}
\begin{align}
h_1'' + k^2 h_1 &= 0\,,\\
h_2'' + k^2 h_2 + a^4 \frac{m_g^2}{2} \, h_2 &= 0\,.
\end{align}
\end{subequations}
with $m_g^2 = y_* m^2\, \Gamma_* $. By inverting \eqref{eq:diagonaliseEOM}, we obtain the composition of eigenstates of matter basis gravitons in terms of the mass eigenstates,
\begin{subequations}
\label{eq:massrotation}
\begin{align}
h(k,t) &= \cos\theta \, h_1 + \sin\theta  \, h_2\,,  \\ 
\tilde{h}(k,t) &= \sin\theta \, h_1 - \cos\theta  \, h_2 \,.
\end{align}
\end{subequations}
Assuming the GW waveform can be modeled as a wave packet, it is sufficient to know the plane wave solution to the equations of motion; the traveling wave packets are then superpositions of plane waves. The GWs are generated via the coupling of $h$ to matter, and subsequently propagate as the mass eigenstates $h_1$ and $h_2$. These will decohere if the flight length exceeds the coherence length~\cite{PhysRevLett.119.111101}
\begin{equation}
\label{eq:coherence_length}
	L_\text{coh} \approx 0.1\, \mathrm{s}\ \frac{ 2 E^2}{m_g^2} = \left( \frac{10^{-21}\, \mathrm{eV}}{m_g}\right)^2  \mathrm{Gpc}\,,
\end{equation}
e.g.~for a plane wave of $E=25$ hz and distance 100 Mpc, if the mass exceeds $m_g~\gtrsim~6~\times~10^{-22}$~eV. The plane wave solutions of \eqref{eq:eom_mass_eigenstates} are
\begin{subequations}
\begin{align}
h_1(k,t) &\propto \cos (k \,t) \,,\\
h_2(k,t) &\propto \cos \Big(\sqrt{k^2+m_g^2} \,t\Big) \approx \cos \Big( \big[k+\frac{m_g^2}{2k}\big] t \Big)\,,
\end{align}
\end{subequations}
where the massive mode propagates as two superimposed oscillations, the plane wave frequency $\omega_0 \equiv k$ and the modulation frequency $\delta \omega \equiv \frac{m_g^2}{2 k}$, valid if $\omega_0 \gg \delta \omega$. By means of Eq.~\eqref{eq:massrotation}, we can now go back to the matter coupling basis and study oscillations, as was done in~\cite{PhysRevLett.119.111101}. However, we now pursue a different path.

\subsection{Decoherence}

In the decoherent regime, these two waves propagate completely independently. The detector response is determined by their overlap with the physical metric, given by \eqref{eq:massrotation}:
\begin{subequations}
\begin{align}
h_a(k,t) &= \cos^2\theta \, \cos (k t)\,,\\
h_b(k,t) &= \sin^2\theta \,  \cos \big((k+\frac{m_g^2}{2k})\,t \big)\,,
\end{align}
\end{subequations}
where we have normalised such that $h(k,0)= 1$ $(h'(k,0) = 0)$ and $\tilde{h}(k,0)=0$. We average out the fast oscillations by integrating over their period $T_0=2\pi /k$ and obtain the suppression factors of the decohered graviton wave packets:
\begin{equation}
\label{eq:supp_decoh}
\langle h_a \rangle  =   \cos^2 \theta\,, \qquad \langle h_b \rangle =   \sin^2 \theta\,.
\end{equation}
Note that this is different to the case discussed in \cite{PhysRevLett.119.111101}, where the interference between $h_1$ and $h_2$ causes a time-dependent modulation of the amplitude.

The results of this section are the suppression factors~\eqref{eq:supp_decoh} of a GW wave packet. This is only valid in the parameter region where the GWs decohere, i.e., when the luminosity distance of the event exceeds the coherence length~\eqref{eq:coherence_length} for all frequency modes. If the mixing angle $\theta$ is non-zero (which recovers GR) and not equal to $\pi/2$ (corresponding to pure massive gravity), a detector of GWs will see two events of approximately the same waveform separated in time, but with their amplitudes rescaled according to~\eqref{eq:supp_decoh}. Note that the waveform corresponding to the massive mode obeys a frequency dependent dispersion in time.

This allows us to probe the parameter space of bigravity where one has a larger mass $m_g$ than relevant for GW oscillations, and small angle $\theta$. This is the subject of the following section.

\section{Phenomenology and Results }\label{sec:results}

\begin{figure*}[t]
\begin{center}
	\subfloat[Merger rate with current uncertainty]{\label{fig:ratio_current}\includegraphics[width=.46\textwidth]{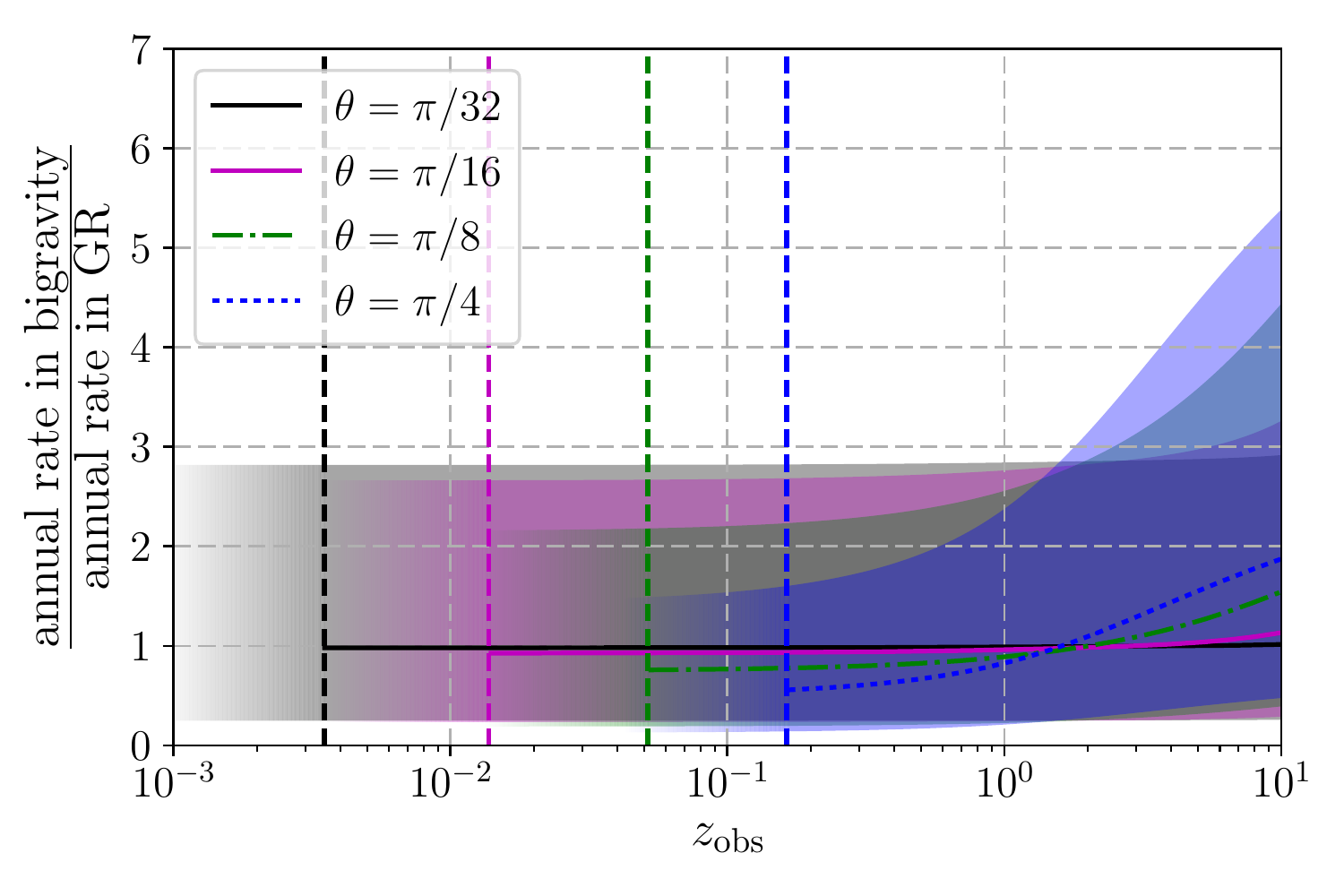}}
    \hspace{3mm}
	\subfloat[Merger rate with projected uncertainty]{\label{fig:ratio_prospects}\includegraphics[width=.46\textwidth]{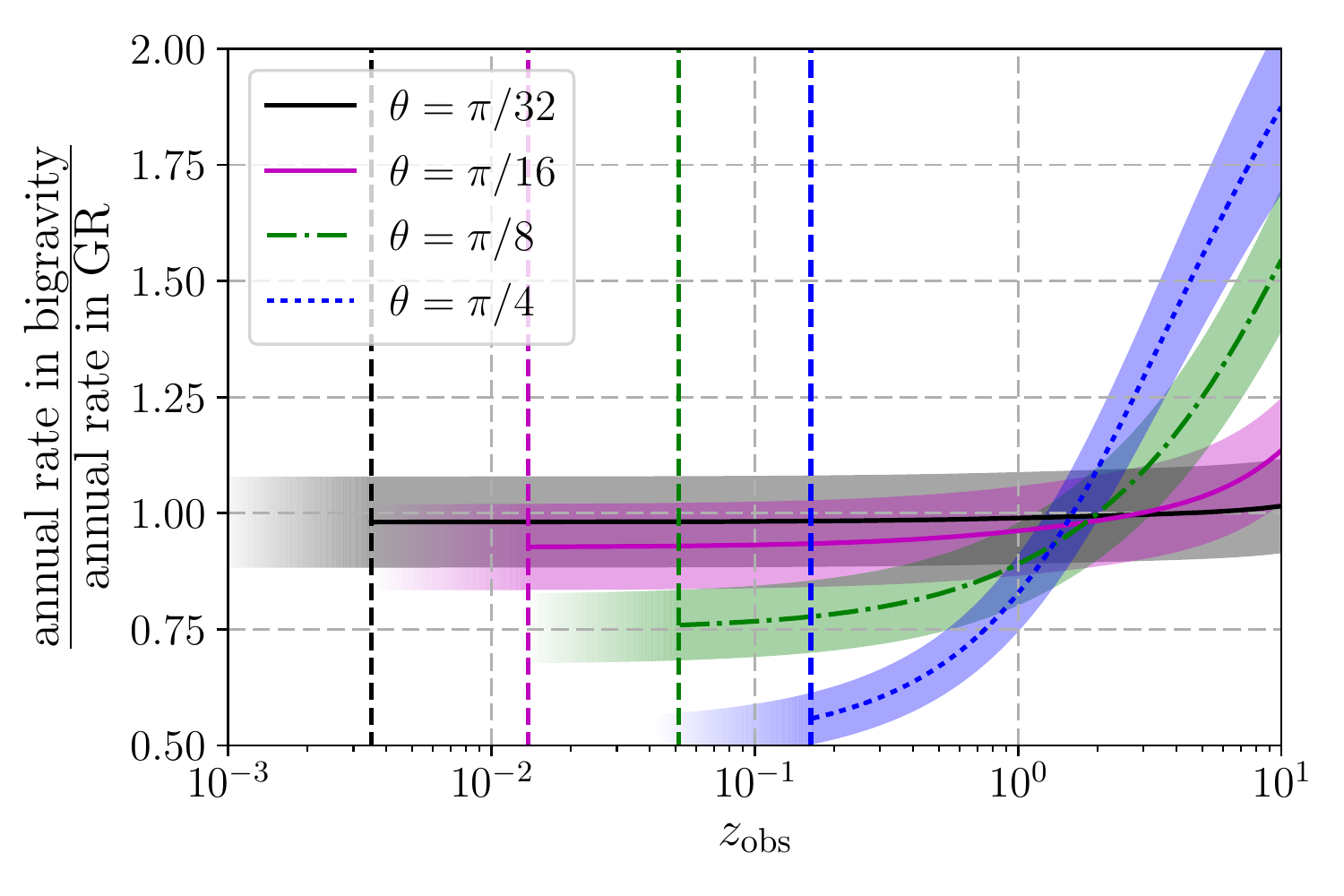}}
\end{center}
\caption{\label{fig:ratio}\emph{Independent events}. Annual rates of BBH merger events for different observed redshifts $z_\text{obs}$. Bigravity in the decohered regime shifts the number of events towards higher $z_\text{obs}$. Rates are normalised to the GR expectation for different mixing angles, based on the BBH merger rate measured to be $R=55^{\,+103}_{\,-41}$\,(Gpc$^3$ yr)$^{-1}$ and the projected rate with $10\%$ accuracy.~\cite{PhysRevX.6.041015,PhysRevD.95.103010} The rates are calculated using $m_g=m_\text{bound} \, \sin\theta$ with the upper bound $m_\text{bound}=7.2~\times~10^{-23}\eV$. Coloured regions indicate the errors and the fading the crossing from the coherent regime to decoherence. At present, all values of the mixing angles are consistent with 1, i.e.~the GR prediction.}
\end{figure*}

In current GW detection measurements (such as the LIGO/Virgo network), the luminosity distance,
\begin{equation}
	d_L (z) = c\, (1+z) \int H(z')^{-1} \mathrm{d}z'\,,
\end{equation}
of a BBH merger event is inferred solely from the amplitude of the strain, as it scales as $h \propto 1/d_L$ \cite{PhysRevLett.116.241102}. In the decoherence regime of bigravity, the strain will suffer a frequency-independent suppression with respect to GR. Without any other means to measure the distance, an event at actual redshift~$z$ will therefore be misinterpreted to stem from an observed redshift $z_\text{obs}$, with $z_\text{obs} > z$. Additionally, the secondary gravitational wave with rescaled amplitude could be seen. We assume in the following the physical graviton to be composed mostly of the massless mode, i.e.~small $\theta$; this means that the wave corresponding to the massless graviton arrives with $z_\text{obs}^{m=0} \gtrsim z$; the subsequent strain of the massive graviton will be interpreted as $z_\text{obs}^{m\neq 0} > z_\text{obs}^{m=0}$. If measurements are sensitive to both signals, one can look for echoed signals arriving shortly after the primary GW signal. Otherwise, a reduction and/or enhancement of the merger rates is expected, i.e.~when one signal isn't seen or both are seen but interpreted as independent events. We will now discuss the two possible interpretations of decoherence in more detail.

\smallskip

\paragraph{Interpretation as independent events}
Let us first take the point of view that the second event is not identified to originate from the same source, or is not seen at all. In this case we will observe a different merger rate than expected in pure GR. Our procedure to quantify this statement is as follows.

First, we must express the number of events per year expected at a given redshift. Following Ref.~\cite{PhysRevD.95.103010}, we define the differential BBH merger rate
\begin{equation}
\label{eq:differential_rate}
\frac{dN}{dz} = 4 \pi \,R \,\chi(z)^2 (1+z)^{(R_b-1)} \frac{c}{H(z)}\,,
\end{equation}
where $\chi(z)$ is the comoving distance, $R_b$ parametrises the redshift-dependence of the merger rate, apart from the expansion of the Universe, and $R$ is the (constant) BBH merger rate density. The above reference finds a best-fit point near $R_b=2$, which we will assume in the following. Note that the fit cannot accommodate redshifts $z \gtrsim 10$.~\cite{PhysRevD.95.103010}

The rate density of BBH mergers can be estimated from the events observed by LIGO/Virgo during the O1 run~\cite{PhysRevX.6.041015}  to be $R = 55^{\,+103}_{\,-41}$\,(Gpc$^3$ yr)$^{-1}$. We expect the rather large errors of this number to decrease significantly in the near future, as LIGO/Virgo collect more data; in~\cite{PhysRevD.95.103010}, it is shown that the merger rate may improve up to $<10\%$ accuracy within a few years of advanced LIGO measurements at design sensitivity.

Using \eqref{eq:differential_rate} we estimate the rate of observable events. Recent GW detections have reached distances about $z~\approx~0.1$. Taking this as the current experimental sensitivity translates to $20_{-15}^{+37}$ observable events per year. This may increase by a factor $10^3$ once advanced LIGO reaches its design sensitivity up to $z\approx 1$ \cite{PhysRevD.93.112004}.

Requiring that the luminosity distance exceeds the coherence length \eqref{eq:coherence_length} leads to a lower bound on~$m_g$. On the other hand, graviton mass bounds from other observations impose the condition $m_g\, \sin \theta \leq m_\text{bound}$, where $m_\text{bound}$ is obtained e.g.~from tests of a modified dispersion relation or modified Newtonian potential of gravity. Combining these bounds, we obtain a minimum distance between BBH merger and observer for decoherence,
\begin{equation}\label{eq:minDist}
\begin{split}
&d_L \gtrsim \frac{2\, \sigma_x \,c \,E^2 }{m_\text{bound}^2} \sin^2\theta \\
=&\, 1.62  \text{ Gpc}  \left(\frac{E}{100 \text{ hz}}\right)^2 \hspace{-.1cm} \frac{\sigma_x}{0.1 \text{ s}} \left(\frac{7.2 \times 10^{-23} \text{ eV}}{m_\text{bound}}\right)^2  \sin^2\theta \,.
\end{split}
\end{equation}
Note that this condition is necessary but not sufficient for decoherence to occur. We employ the bound $m_\text{bound}~=~7.2~\times~10^{-23} \eV$ from solar system tests~\cite{deRham:2016nuf}, which is the most stringent, model-independent bound available.

\begin{figure*}[t]
\begin{center}
\subfloat[Suppression factor and bound from GW170817]{\label{fig:echo_GW170817}
\includegraphics[width=.48\textwidth]{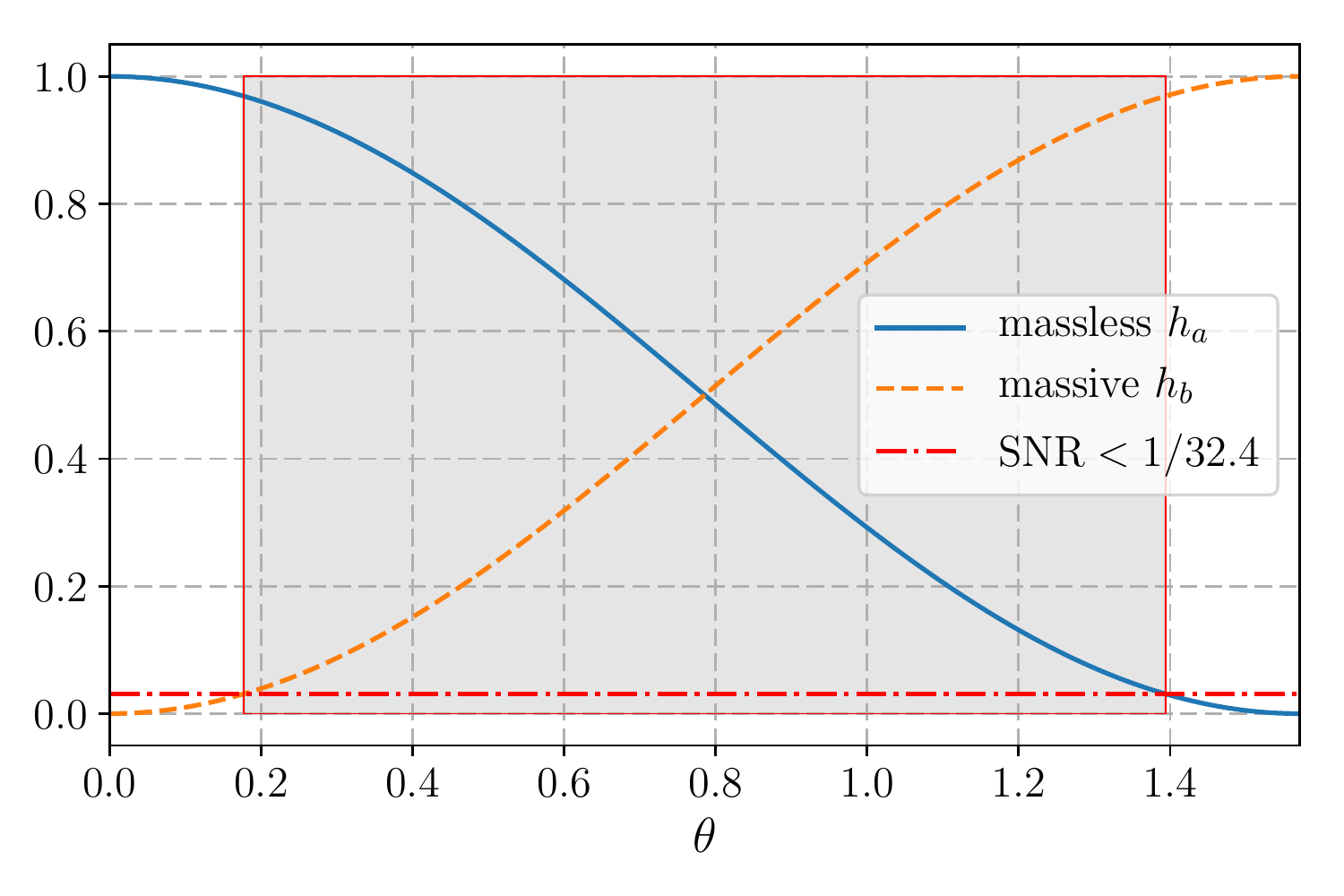}
}
\hfill
\subfloat[Summary of echo event constraints]{\label{fig:exclusion_plot}
\includegraphics[width=.48\textwidth]{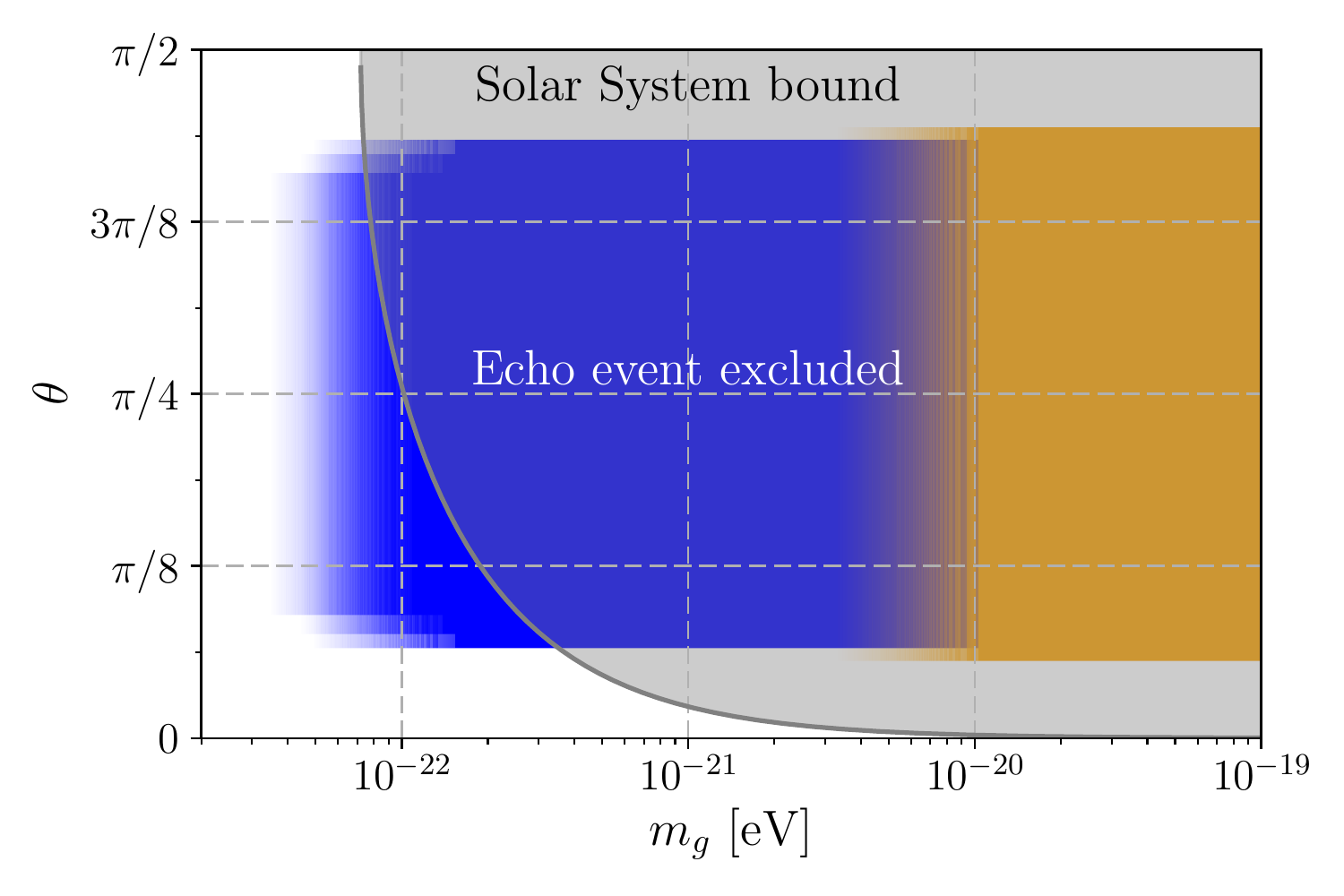}
}
\end{center}
\caption{\label{fig:dec_signals}\emph{Echoed Events}. \emph{Left}: Suppression factor~\eqref{eq:supp_decoh} of a traveling GW for varying mixing angle $\theta$ in the decohered parameter regime. Requiring the secondary waveform to be lost in detector noise excludes a large parameter range (shaded area). \emph{Right}: Excluded parameter range for all available GW events published by LIGO/Virgo (blue: black hole mergers, orange: neutron star merger).~\cite{PhysRevLett.116.241102,Abbott:2017vtc,Abbott:2017gyy,Abbott:2017oio,PhysRevLett.119.161101,2016PhRvL.116x1103A} The fading indicates the onset of the decoherence regime ($L_\text{coh} < d_L < 10 \times L_\text{coh}$). Also shown is the bound $m_g~\leq \sin\theta \times 7.2\times10^{-23} \eV$ from solar system tests~\cite{deRham:2016nuf}.}
\end{figure*}

We now consider a BBH merger event at true redshift~$z$. Given the suppression factor for the physical metric $g$, we determine the interpreted redshift $z_\text{obs}$ by solving the equation
\begin{equation}
	\cos^2 \theta / d_L(z) = 1 / d_L(z_\text{obs})\, ,
\end{equation}
which is found from the fact that $h \propto 1/d_L(z)$. In the next step, we calculate the observed merger rate $\left.\frac{\mathrm{d}N}{\mathrm{d}z}\right|_{z = z_\text{obs}}$ for a given redshift. The results are summarised in Fig.~\ref{fig:ratio}. Note two important features: first, we have also included the (current and projected) errors on the merger rate density as estimated by LIGO/Virgo. This is indicated by the coloured bands in Fig.~\ref{fig:ratio}. Secondly, we have included the effect that decoherence requires events to occur at a certain distance/redshift $z_\text{min}$ which is a function of the mixing angle, cf.~Eq.~\eqref{eq:minDist}. This is represented by the dashed vertical lines. As it is an order-of-magnitude criterion, the bands fade out below~$z_\text{min}$.

Normalised to the rate predicted in GR, we find that small redshift events are more scarce than in GR (ratio~$<1$), while large redshift events appear to be more abundant (ratio~$>1$).
Clearly, the observation of an echoed waveform would be a striking signature for observations, and one would naturally assume the two consecutive detections to be correlated. However, the current analysis equally applies to the low-$z$ range: if the echoed signal is lost in detector noise, one may still observe \textit{too few} events compared to the prediction at lower~$z$. In other words, the echo event's observed redshift is simply beyond the detector reach. This is the power of the present approach: we do not require a correlation of two events and/or a measured time difference between them (which would depend on the parameters of the binary system). For a desired sensitivity, a sufficiently large number of events at different redshifts has to be observed, in order to draw conclusions on the merger rate density and thereby on the  bigravity parameter space.

In conclusion, no restrictions arise from this analysis at present, since all mixing angles are consistent with the annual merger rate predicted in GR (ratios consistent with $1$). However, Fig.~\ref{fig:ratio_prospects} shows that in the future, with a 10\% uncertainty on the merger rate, one can clearly distinguish the cases with large mixing angles $\theta \gtrsim \pi/16$ from the GR predicted merger rate.

\smallskip

\paragraph{Interpretation as echoed events}

If the parameters of bigravity lie in the decoherence regime for astrophysical distances, an initial wave packet of a GW event will split into two waveforms. The waveform corresponding to the massless modes arrives first and is simply the GR waveform rescaled by $\cos^2\theta$. The waveform composed of massive modes is distorted due to the frequency-dependence of the time delay $\Delta t \propto 1/E^2$. The LIGO-Virgo detector network should then see a distinctive signal of two separated events with a time separation of the order of seconds.

The rescaling of the separated wave packets is given by Eq.~\eqref{eq:supp_decoh}. Requiring that the secondary signal (due to the massive mode, if bigravity is mostly massless, or vice versa) is not observed due to the finite signal-to-noise ratio (SNR) of the detectors, we can set a limit on the bigravity mixing angle $\theta$, assuming that $m_g$ is such that the wave packets are fully separated. The highest SNR so far achieved stems from the neutron star merger observation GW170817, with a combined SNR of 32.4.~\cite{TheLIGOScientific:2017qsa} For the mostly-massless scenario, this restricts $\theta \lesssim 0.18 \approx \frac{\pi}{16}$, while for a mostly-massive graviton $\theta \gtrsim 1.39 \approx \frac{7\pi}{16}$ is required, see Fig.~\ref{fig:echo_GW170817}.

Recall however that the bound only applies if $d_L~>~L_\text{coh}$ is satisfied. Under this constraint, the~bounds on the parameter space $(\theta,m_g)$ are summarised in Fig.~\ref{fig:exclusion_plot} for all GW events observed so far. Once more, the fading indicates the transition into the decoherence regime. We find that even though the most stringent bound on $\theta$ is obtained from the high SNR of GW170817 (orange), the large temporal width~$\mathcal{O}(60\,$s) of this signal requires a large $m_g$ in order for the wave packets to separate sufficiently; the bound is thus overlapped by the local gravity bound. Better results are achieved from the BBH merger observations, which are of shorter durations $\mathcal{O}(1\,$s) (blue).

Finally, we comment on the search for said echoed signals with LIGO/Virgo. The time separation of the two events is given by \eqref{eq:coherence_length}, requiring the event separation $\Delta t$ to be of order of the signal width $\sigma_x$. For currently probed distances and graviton masses in the decohered regime which are not excluded by observations, this translates into $\Delta t= \mathcal{O}(0.1\div1\,$s).

We thus propose a template search: for every detected GW event, one should investigate the region of a few seconds about the triggering signal and look for an echoed event. The waveform of the secondary strain is fully determined by the triggering waveform and the dispersion relation of massive gravitons. A complete analysis of the modified observational signature is however beyond the reach of this publication.

Finally, we wish to point out that such echo signals have recently been proposed in the context of exotic compact objects, which mimic black holes at large distances, but modify the near-horizon physics, see e.g.~\cite{Mark:2017dnq,Cardoso:2016oxy} and references therein for realisations. In these cases the horizon of the black hole is replaced with a reflective surface, such that part of the GW signal is reflected. In contrast to our proposal, these echo signals would not only entail one (or $N-1$ in the case of $N$-metric gravity~\cite{Hinterbichler:2012cn}), but infinitely many echoes, with decreasing amplitude. Furthermore, if an echo is created due to decoherence, the time difference to the initial signal must grow with the distance of the merger.

However, a recent claim that such echoes are in fact seen by LIGO/Virgo~\cite{Abedi:2016hgu,Abedi:2017isz,Conklin:2017lwb} is under debate~\cite{Ashton:2016xff,Westerweck:2017hus} in the literature. It is therefore too early to make any claims in this direction. We may tentatively note, however, that the $\Delta t$ obtained in \cite{Conklin:2017lwb} fits the above scenario for $m_g~=~10^{-22}\div 10^{-23}\,$eV.

\section{Conclusions}\label{sec:Conclusions}

We have discussed the decoherence regime of gravitational wave oscillations in the framework of our previous analysis in Ref.~\cite{PhysRevLett.119.111101}. In short, the two propagating wave packets, massive and massless, will no longer overlap spatially in this regime, and two signals can in principle be seen in the GW detector. We find that two possible interpretations are feasible. One, where the second event is interpreted as an independent event, and another, where the second event is treated as an echo of the first event. 

In the former case, we find no further constraints on the parameter space of bigravity, assuming otherwise only solar system tests. With more events and better precision, this will significantly improve over the forthcoming years when more events are available and the uncertainty on the annual merger rate decreases. In the latter interpretation, on the other hand, we find a phenomenological bound $\theta \lesssim \pi/16$ for the mostly-massless and $\theta \gtrsim 7\pi/16$ for the mostly-massive scenario, assuming a large enough~$m_g$, such that the corresponding wave packets are non-overlapping.

Finally, we comment on the possibility of having both GW and optical signals, as was the case for the neutron star binary merger GW170817.~\cite{TheLIGOScientific:2017qsa} Such an event is in principal very appealing because it could allows a direct comparison of the speed of the GW and the optical signal. However, at present, no reliable estimate for the different emission times is available, and thus only model dependent constraints arise: It is commonly assumed that the optical signal in the form of a gamma-ray burst (GRB) is emitted within a few $100\,$ms of the GW chirp. Albeit, for GW170817, the GRB was observed $1.7\,s$ \emph{after} the GW chirp. Instead of a modified dispersion relation, this could simply indicate that the optical signal was delayed by ejected material, see~\cite{Ciolfi:2014yla} for a mechanism that delays the GRB by more than $\mathcal{O}(10^3\,s)$. The present analysis stays agnostic to such model-dependent production/emission mechanisms and thus puts more general bounds on the parameter space of bigravity based on the propagation of the GWs only.

We conclude with the proposal of a template search, where one searches the observational data for a secondary waveform of a detected GW event, separated by up to a few seconds. The triggering event can be searched for with the current analysis methods, as it is only a rescaled version of the strain predicted by GR. This provides a clear signature for bigravity in the decoherent regime, and, while current claims of observations are under debate, the search method has been proposed in the literature and is readily applied to our scenario.

\acknowledgements
KM would like to thank the MPIK for hospitality during which part of this project was realised. We thank E.~Most, E.~Trincherini, R.~Contino, and B.~Patricelli for useful discussions. MP is supported by IMPRS-PTFS.

\bibliography{literature}

\end{document}